\documentstyle[12pt]{article}
\renewcommand{\baselinestretch}{1.3}       %
\newcommand{\beq}{\begin{equation}}
\newcommand{\eeq}{\end{equation}}
\newcommand{\ep}{\epsilon}

\newcommand{\bea}{\begin{eqnarray}}
\newcommand{\eea}{\end{eqnarray}}

\newcommand{\mbx}[1]{\mbox{\small$#1$}}
\newcommand{\hmbx}[1]{\hspace*{-3.5mm}\mbox{\small$#1$}}

\title{Five-loop renormalization 
group functions of  ${O}(n)$-symmetric $\phi^4$-theory 
and $\ep$-expansions of critical exponents
up to $\ep^5$ } 

\vspace{2.5cm}

\author{ H.\ Kleinert, J.\ Neu and V.\ Schulte-Frohlinde\thanks
                {Work supported in part by
                 Deutsche Forschungsgemeinschaft
                 under grant no.\ Kl.~256.} \vspace{-2mm} \\
        {
        \small
        Institut f\"{u}r Theoretische Physik}\vspace{-4mm} \\
        {
        \small
        Freie Universit\"{a}t Berlin} \vspace{-4mm}\\
        {
        \small
        Arnimallee 14} \vspace{-4mm}\\
        {
        \small
        D - 1000 Berlin 33} \vspace{5mm}\\
K.G.\ Chetyrkin and S.A.\ Larin \vspace{-2mm}\\
{
\small
Institute for Nuclear Research of the}\vspace{-4mm}\\
{
\small
Academy of Sciences of the USSR }\vspace{-4mm}\\
{\small
60th October Anniversary prospect 7a
}\vspace{-4mm}\\
{\small
Moscow 117312, USSR }\vspace{-4mm}\\
       }
\date{
}

\begin{document}
\vspace{1cm}
\maketitle

\begin{abstract}
\noindent
Motivated by the discovery of errors in six of the 135 diagrams
in the published
five-loop
expansions of the $\beta$-function and the
anomalous dimensions of the ${ O}(n)$-symmetric
$\phi^4$-theory in $D=4-\ep$ dimensions
we present the results of a full analytic reevaluation
of all diagrams.
The divergences are removed
by minimal subtraction
and $\ep$-expansions are given for the
critical exponents $\eta$, $\nu$, and $\omega$ up to order $\epsilon ^5$.

\end{abstract}
\renewcommand{\baselinestretch}{1.6}

\newpage

\noindent
1) During the last two decades, much effort has been invested into 
studying the
scalar quantum field theory  with $\phi^4$-interaction.
On the one hand, such a theory describes correctly many
experimentally observable features of critical phenomena.
Field theoretic renormalization group
techniques~\cite{bogshir} in $D=4-\ep$ dimensions~\cite{wilson,fisher,brezin}
combined with Borel
resummation methods of the resulting
$\ep$-expansions~\cite{zj} led to extremely
accurate determinations of the critical exponents
of all ${O}(n)$ universality
classes. The latter requires the knowledge of the
asymptotic behaviour of perturbation series in four dimensions
which is completely known in this theory \cite{lip}.
Apart from such important applications,
the $\phi^4$-theory, being the
simplest renormalizable quantum field theory in the four dimensional
space-time, has been an ideal ground for testing
new methods of calculating
Feynman diagrams and for studying the structure of perturbation theory.

The RG functions of the $\phi^4$-theory were first calculated
analytically in four dimensions using dimensional regularization \cite{dim}
and the minimal subtraction (MS) scheme~\cite{ms}
in the three- and four-loop approximations in Refs. \cite{nickel}
and \cite{vkt}.
The
critical exponents  were obtained as
$\ep$-expansions~\cite{fisher} up to terms of order
$\ep^3$ and $\ep^4$.

The five-loop anomalous dimension of the field $\phi$ and
the associated critical exponent $\eta$ to order $\ep^5$ 
were
determined analytically in \cite{ckt}.
The five-loop
$\beta$-function and the anomalous dimension of the mass  were
given in Ref.~\cite{cglt}.
However, three
of the 124 four-point
 diagrams contributing to
the $\beta$-function at the five-loop level
could be evaluated only numerically.
The analytic calculation of the
$\beta$-function
was finally completed in \cite{kaz}. The 
ensuing
$\ep$-expansions
for the critical exponents were obtained up to order $\ep^5$ in 
\cite{ep5}.

Intending further applications,
the Berlin group of the authors
undertook an
independent recalculation of the perturbation series
of Refs.~\cite{ckt,cglt}, using the same techniques,
and discovered errors in six of the 135 diagrams.
This meant that
the subsequent results of~\cite{kaz,ep5}
were also incorrect.
When visiting the Moscow group the errors were confirmed
and we can now jointly report all expansions in the correct form.

\noindent
2)
We consider the $O(n)$-symmetric theory of $n$ real scalar
 fields $\phi^a$ ($a=1,2,\ldots,n$) with the Lagrangian
\begin{equation}
\label{lagr}
L=\frac{1}{2} \partial_{\mu}\phi^a\partial_{\mu}\phi^a+\frac{m_B^2}{2}
\phi^a\phi^a+\frac{16\pi^2}{4!}g_B(\phi^a\phi^a)^2
\end{equation}
in an euclidean space with $D=4-\ep$ dimensions.
The bare (unrenormalized) coupling constant $g_{\rm B}$ and mass $m_{\rm B}$
are expressed
via renormalized ones as
\begin{equation}
\label{bare}
g_{\rm B}=\mu^{\ep}Z_gg=\mu^{\ep}\frac{Z_4}{(Z_2)^2}g,
~~~~m_{\rm B}^2=Z_{m^2}m^2=\frac{Z_{\phi^2}}{Z_2}m^2
\ .
\end{equation}
Here $\mu$ is the unit of mass in dimensional regularization and
$Z_4$, $Z_2$,
$Z_{m^2}$ are the renormalization constants of the vertex function,
propagator and mass, respectively, with $Z_{\phi^2}$ being the
renormalization constant of the
two-point function obtained from the propagator by the insertion, in all
possible ways, of the vertex ($-\phi^2$)~\cite{vkt}.
In the MS-scheme the renormalization constants do not depend on dimensional
parameters and are expressible as series in  $1/\ep$ with
purely $g$-dependent coefficients:
\begin{equation}
\label{z}
Z_i=1+\sum_{k=1}^{\infty} \frac{Z_{i,k}(g)}{\ep^k}
\ .
\end{equation}
The $\beta$-function and the anomalous dimensions entering the RG equations
are expressed in the standard way as follows:
\begin{equation}
\label{rgdef1}
\beta(g)=\frac{\ep}{2}\,g
+\left.\frac{d~g}{d~{\ln}\,\mu^2}\right|_{g_{\rm B}}=
\frac{1}{2}g\frac{\partial~Z_{g,1}}{\partial~g}
\ ,
\end{equation}
\beq \gamma _{m}=\left.\frac{d~{\ln}\,m}{d~{\ln}\,\mu}\right|_{g_{\rm B}}
=-\frac{d~{\ln}\,Z_{m^2}}{d~{\ln}\,\mu^2}
=\frac{1}{2}g\frac{\partial~Z_{m^2,1}}{\partial~g}
\label{rgdef2}
\ ,
\eeq
\beq
\gamma_i(g)=\left.\frac{d~{\ln}\,Z_i}{d~{\ln}\,\mu^2}\right|_{g_{\rm B}}
=-\frac{1}{2}g\frac{\partial~Z_{i,1}}{\partial~g}
\ ,~~~i=2,4,\phi^2
\ .
\label{rgdef3}
\eeq
We also use the relations
\begin{equation}
\label{useful}
\beta (g)=g[2\gamma_2(g)-\gamma_4(g)]
\ ,~~~\gamma_{m}(g)=\gamma_2(g)-\gamma_{\phi^2}(g),
\end{equation}
which follow from the relations between renormalization constants
implied by (\ref{bare})
and are useful for the calculations of $\beta (g)$ and $\gamma_{m}(g)$.

To determine all RG functions up to five loops we calculate the
five-loop approximation to the three constants $Z_2$, $Z_4$ and
$Z_{\phi^2}$.  The constant $Z_2$ contains the counterterms of
the 11 five-loop propagator diagrams.  Two of them were calculated
erroneously in Ref.~\cite{ckt}. The constant $Z_4$ receives contributions
from 124 vertex diagrams.
Of these diagrams, 90 contribute to
$Z_{\phi^2}$ after appropriate changes of combinatorial factors.
Four of the 124 counterterms were calculated erroneously in
Ref.~\cite{cglt}.

In the present paper we have used the same methods 
as in the previous works~\cite{ckt,cglt}
to
calculate the counterterms from the dimensionally regularized
Feynman integrals, namely,
the
method of infrared rearrangement \cite{vlad}, the
Gegenbauer polynomial $x$-space
technique~(GPXT)~\cite{tpgx},
the integration-by-parts algorithm \cite{ibp}, and the
$R^*$-operation~\cite{rstar}. 
Three diagrams 
were calculated analytically
first in \cite{kaz}
by using the so-called method of uniqueness, later the same results
were obtained for them by using the Gegenbauer polynomials in $x$-space
together
with several
non-trivial tricks \cite{brod}.
A detailed description of the calculations 
including the diagramwise results
will be presented in a separate publication.

The analytic results of our recalculation of the five-loop approximations
to the RG functions $\beta(g)$, $\gamma _2(g)$ and $\gamma _m(g)$
are [$\zeta(n)$ is the Riemann $\zeta$-function]:
\bea 
\lefteqn{\mbx{\beta(g)
\ =\     \frac{g^2}{6}\bigl[n+8\bigr]
\ -\ \frac{g^3}{6}\bigl[3n+14\bigr]} }
\nonumber \\
&\mbx{+\frac{g^4}{432} }
 &\hmbx{ \bigl[33n^2+922n+2960\,
       +\,\zeta(3)\cdot96(5n+22)\bigr] }
\nonumber \\
&\mbx{-\frac{g^5}{7776} }
 &\hmbx{ \bigl[-5n^3+6320n^2+80456n+196648 }
\nonumber \\
&
 &\hmbx{+\,\zeta(3)\cdot96(63n^2 + 764n + 2332) }
\nonumber \\
&
 &\hmbx{-\,\zeta(4)\cdot288(5n+22)(n+8) }
\nonumber \\
&
 &\hmbx{+\,\zeta(5)\cdot1920(2n^2 + 55n + 186)\bigr] }
\nonumber \\
&\mbx{+\frac{g^6}{124416} }
 &\hmbx{ \bigl[13n^4+12578n^3+808496n^2 +6646336n+13177344 }
\nonumber \\
&
 &\hmbx{+\,\zeta(3)\cdot16(-9n^4+1248n^3+67640n^2+552280n+1314336) }
\nonumber \\
&
 &\hmbx{+\,\zeta^2(3)\cdot768(-6n^3-59n^2+446n+3264) }
\nonumber \\
&
 &\hmbx{-\,\zeta(4)\cdot288(63n^3+1388n^2+9532n+21120) }
\nonumber \\
&
 &\hmbx{+\,\zeta(5)\cdot256(305n^3+7466n^2+66986n+165084) }
\nonumber \\
&
 &\hmbx{-\,\zeta(6)(n+8)\cdot9600(2n^2+55n+186) }
\nonumber \\
&
 &\hmbx{+\,\zeta(7)\cdot112896(14n^2 + 189n + 526)\bigr] }
\ ,
\label{analbeta}
\eea 
\bea 
\lefteqn{\!\!\!\!\!\!\!\!\!\!\!\!\!\!\!\!\!\!\!
\!\!\!\!\!\!\!\!\!\!\!\!\!\!\!\!\!\!\!\!\!\!\mbx{ \gamma_2(g)
\ =\     \frac{g^2}{36}(n+2)
\ -\ \frac{g^3}{432}(n+2)\bigl[n+8\bigr]} }
\nonumber \\
&\!\!\!\!\!\!\!\!\!\!\!\!\!\!\!\!\mbx{+\frac{g^4}{5184}(n+2) }
 &\hmbx{\bigl[5(-n^2+18n+100)\bigr] }
\nonumber \\
&\mbx{-\frac{g^5}{186624}(n+2) }
 &\hmbx{\bigl[39n^3+296n^2+22752n+77056 }
\nonumber \\
&
 &\hmbx{-\,\zeta(3)\cdot48(n^3-6n^2+64n+184) }
\nonumber \\
&
 &\hmbx{+\,\zeta(4)\cdot1152(5n + 22)\bigr] }
\ ,
\label{analgam2}
\eea 
\bea 
\lefteqn{\!\!\!\!\!\!\!\!\!\!\!\!\!\!\!\!\mbx{\gamma_{m}(g)
\ =\     \frac{g}{6}(n+2)
\ -\ \frac{g^2}{36}(n+2)\bigl[5\bigr]
\ +\ \frac{g^3}{72}(n+2)\bigl[5n+37\bigr]} }
\nonumber \\
&\mbx{-\frac{g^4}{15552}(n+2) }
 &\hmbx{\bigl[-n^2+7578n+31060 }
\nonumber \\
&
 &\hmbx{+\,\zeta(3)\cdot48(3n^2+10n+68) }
\nonumber \\
&
 &\hmbx{+\,\zeta(4)\cdot288(5n+22)\bigr] }
\nonumber \\
&\mbx{+\frac{g^5}{373248}(n+2) }
 &\hmbx{\bigl[21n^3+45254n^2+1077120n+3166528 }
\nonumber \\
&
 &\hmbx{+\,\zeta(3)\cdot48(17n^3+940n^2+8208n+31848) }
\nonumber \\
&
 &\hmbx{-\,\zeta^2(3)\cdot768(2n^2+145n+582) }
\nonumber \\
&
 &\hmbx{+\,\zeta(4)\cdot288(-3n^3+29n^2+816n+2668) }
\nonumber \\
&
 &\hmbx{+\,\zeta(5)\cdot768(-5n^2+14n+72) }
\nonumber \\
&
 &\hmbx{+\,\zeta(6)\cdot9600(2n^2+55n+186)\bigr] }
\ .
\label{analgamm2}
\eea 
For $n=1$ the series have the numerical
form:
\begin{equation}
\begin{array}{c}
\label{numbeta}
\beta(g)=
 1.5\,g^2
-2.833\,g^3
+ 16.27\,g^4
- 135.8\,g^5
+1424.2841\,g^6
\ ,
\end{array}
\end{equation}
\begin{equation}
\begin{array}{c}
\label{numgam2}
\gamma_2=
   0.0833\,g^2
- 0.0625\,g^3
 + 0.3385\,g^4
-1.9256\,g^5
\ ,
\end{array}
\end{equation}
\begin{equation}
\begin{array}{c}
\gamma_{m}=
 0.5\,g
- 0.4167\,g^2
  + 1.75\,g^3
 - 9.978\,g^4
+ 75.3778\,g^5
\ .
\end{array}
\end{equation}
Note that the five-loop coefficients have changed by about 0.3
\% for the  $\beta$-function, by about 9 \% for $\gamma_{m}$,
and by a factor of three for $\gamma_2$ in comparison with the
wrong results of Refs.~\cite{ckt,cglt}.

\noindent
3) These RG functions can now be used to calculate the
critical
exponents describing the behaviour of
a statistical system near the critical point
of the second order phase transition \cite{brezin}. At the critical
temperature $T=T_{\rm C}$, the
asymptotic behaviour
of the correlation function for
$|{\bf x}|\rightarrow\infty$ has the form
\begin{equation}
\label{cor}
\Gamma ({\bf x})\sim \frac{1}{{\bf |x|}^{D-2+\eta}}
\ .
\end{equation}
Close to $T_{\rm C}$, the correlation length
behaves for $t=T-T_{\rm C}\rightarrow 0$ as
\begin{equation}
\label{xi}
\xi \sim t^{-\nu} (1+{\rm const}\cdot t^{\omega \nu}+\ldots)
\ .
\end{equation}
The three critical exponents $\eta$, $\nu$ and $\omega$
defined in this
way completely specify the critical behaviour of the system.
All other
exponents can be expressed in terms of these~\cite{brezin}.

The three critical exponents can be determined from
the RG functions of the
$\phi^4$-theory by going to the
infrared-stable fixed point
\begin{equation}
\label{g0}
g=g_0(\ep)=\sum_{k=1}^{\infty} g^{(k)}\ep^k
\end{equation}
which is determined by the condition
($\beta_\ep\equiv \beta - \frac{\ep}{2} g$)
\begin{equation}
\label{zero}
\beta_{\ep}'(g_0)=0,~~~~
\beta_{\ep}(g_0)=\bigl[ \partial \beta_{\ep}(g)/
\partial g \bigr]_{g=g_0} > 0
\ .
\end{equation}

The resulting formulas for the critical exponents are:
\begin{equation}
\label{form}
\begin{array}{c}
\eta=2\gamma_2(g_0)\ ,
~~~1/\nu=2(1-\gamma_{m}(g_0))\ ,
~~~w=2\beta_{\ep}'(g_0)\ ,
\end{array}
\end{equation}
each emerging as an $\ep$-expansion up to order $\ep^5$.
From (\ref{analbeta})-(\ref{analgamm2}) we therefore find:
\bea 
\lefteqn{\!\!\!\!\!\!\!\!\!\!\!\!\!\!\!\!\!\!\!\!\!\!
\mbx{\eta(\ep)
\ =\     \frac{(n+2)\ep^2}{2(n+8)^2}
         \Bigl\{1+\frac{\ep}{4(n + 8)^2}
              \bigl[-n^2+56n+272\bigr]} }
\nonumber \\
&\mbx{-\frac{\ep^2}{16(n+8)^4} }
 &\hmbx{\bigl[5n^4+230n^3-1124n^2-17920n-46144 }
\nonumber \\
&
 &\hmbx{+\,\zeta(3)(n+8)\cdot384(5n+22)\bigr] }
\nonumber  \\
&\mbx{-\frac{\ep^3}{64(n+8)^6} }
 &\hmbx{\bigl[13n^6+946n^5+27620n^4+121472n^3 }
\nonumber \\
&
 &\hmbx{-262528n^2-2912768n-5655552 }
\nonumber \\
&
 &\hmbx{-\,\zeta(3)(n+8)\cdot16\hspace*{-2mm}
         \begin{array}[t]{l}
                      \mbx{(n^5+10n^4+1220n^3-1136n^2}
           \\[-1.5mm] \mbx{-68672n-171264)}
         \end{array} }
\nonumber \\ 
&
 &\hmbx{+\,\zeta(4)(n+8)^3\cdot1152(5n+22) }
\nonumber \\
&
 &\hmbx{-\,\zeta(5)(n+8)^2\cdot5120(2n^2+55n+186)\bigr]\Bigr\} }
\ , 
\label{analcrit1}
\eea 
\bea 
\lefteqn{\mbx{1/\nu(\ep)
\ =\   2+\frac{(n+2)\ep}{n+8}\Bigl\{
       -1
      -\frac{\ep}{2(n+8)^2}\bigl(13n+44\bigr)} }
\nonumber \\
&\mbx{+\frac{\ep^2}{8(n+8)^4} }
 &\hmbx{\bigl[3n^3-452n^2-2672n-5312 }
\nonumber \\
&
 &\hmbx{+\,\zeta(3)(n+8)\cdot96(5n+22)\bigr] }
\nonumber \\
&\mbx{+\frac{\ep^3}{32(n+8)^6} }
 &\hmbx{\bigl[3n^5+398n^4-12900n^3-81552n^2-219968n-357120 }
\nonumber \\
& 
 &\hmbx{+\,\zeta(3)(n+8)\cdot16(3n^4-194n^3+148n^2+9472n+19488) }
\nonumber \\
&
 &\hmbx{+\,\zeta(4)(n+8)^3\cdot288(5n+22) }
\nonumber \\
&
 &\hmbx{-\,\zeta(5)(n+8)^2\cdot1280(2n^2+55n+186)\bigr] }
\nonumber \\
&\mbx{+\frac{\ep^4}{128(n+8)^8} }
 &\hmbx{\bigl[3n^7-1198n^6-27484n^5-1055344n^4 }
\nonumber \\
&
 &\hmbx{-5242112n^3-5256704n^2+6999040n-626688 }
\nonumber \\
&
 &\hmbx{-\,\zeta(3)(n+8)\cdot16\hspace*{-2mm}
        \begin{array}[t]{l}
                     \mbx{(13n^6-310n^5+19004n^4+102400n^3}
          \\[-1.5mm] \mbx{-381536n^2-2792576n-4240640)} 
        \end{array} }
\nonumber \\
&
 &\hmbx{-\,\zeta^2(3)(n+8)^2\cdot1024(2n^4+18n^3+981n^2+6994n+11688)}
\nonumber \\
&
 &\hmbx{+\,\zeta(4)(n+8)^3\cdot48(3n^4-194n^3+148n^2+9472n+19488) }
\nonumber \\
&
 &\hmbx{+\,\zeta(5)(n+8)^2\cdot256(155n^4+3026n^3+989n^2-66018n-130608) }
\nonumber \\
&
 &\hmbx{-\,\zeta(6)(n+8)^4\cdot6400(2n^2+55n+186) }
\nonumber \\
&
 &\hmbx{+\,\zeta(7)(n+8)^3\cdot56448(14n^2+189n+526)\bigr]\Bigr\} }
\ ,
\label{analcrit2}
\eea 
\bea 
\lefteqn{\mbx{\omega(\ep)
\ =\     \ep
        -\frac{\ep^2}{(n+8)^{2}}\bigl[9n+42\bigr]} }
\nonumber \\
&\mbx{+\frac{\ep^3}{4(n+8)^{4}} }
 &\hmbx{\bigl[33n^3+538n^2+4288n+9568 }
\nonumber \\
&
 &\hmbx{+\,\zeta(3)(n+8)\cdot96(5n+22)\bigr] }
\nonumber \\
&\mbx{+\frac{\ep^4}{16(n+8)^{6}} }
 &\hmbx{\bigl[5n^5-1488n^4-46616n^3-419528n^2-1750080n-2599552 }
\nonumber \\
&
 &\hmbx{-\,\zeta(3)(n+8)\cdot96(63n^3+548n^2+1916n+3872) }
\nonumber \\
&
 &\hmbx{+\,\zeta(4)(n+8)^3\cdot288(5n+22) }
\nonumber \\
&
 &\hmbx{-\,\zeta(5)(n+8)^2\cdot1920(2n^2+55n+186) \bigr] }
\nonumber \\
&\mbx{+\frac{\ep^5}{64(n+8)^{8}} }
 &\hmbx{\bigl[13n^7+7196n^6+240328n^5+3760776n^4 }
\nonumber \\
&
 &\hmbx{+38877056n^3+223778048n^2+660389888n+752420864 }
\nonumber \\
&
 &\hmbx{-\,\zeta(3)(n+8)\cdot16\hspace*{-2mm}
         \begin{array}[t]{l}
                       \mbx{(9n^6-1104n^5-11648n^4-243864n^3}
            \\[-1.5mm] \mbx{-2413248n^2-9603328n-14734080)}
          \end{array} }
\nonumber \\
&
 &\hmbx{-\,\zeta^2(3)(n+8)^2\cdot768(6n^4+107n^3+1826n^2+9008n+8736)}
\nonumber \\
&
 &\hmbx{-\,\zeta(4)(n+8)^3\cdot288(63n^3+548n^2+1916n+3872) }
\nonumber \\
&
 &\hmbx{+\,\zeta(5)(n+8)^2\cdot256(305n^4+7386n^3+45654n^2+143212n+226992) }
\nonumber \\
&
 &\hmbx{-\,\zeta(6)(n+8)^4\cdot9600(2n^2+55n+186) }
\nonumber \\
&
 &\hmbx{+\,\zeta(7)(n+8)^3\cdot112896(14n^2+189n+526)\bigr] }
\ .
\label{analcrit3}
\eea 

For $n=1$, these expansions read, numerically,
\beq
\label{numcrit1}
\begin{array}{c}
\eta=
0.01852\,\ep^2
+0.01869\,\ep^3
-0.00833\,\ep^4
+0.02566\,\ep^5
\ , 
\end{array}
\eeq
\beq
\label{numcrit2}
\begin{array}{c}
\frac{1}{\nu}=
 2
- 0.333\,\ep
- 0.1173\,\ep^2
+ 0.1245\,\ep^3
- 0.307\,\ep^4
+ 0.951\,\ep^5
\ , 
\end{array}
\eeq
\beq
\label{numcrit3}
\begin{array}{c}
\omega=
\ep
- 0.630\,\ep^2
+ 1.618\,\ep^3
- 5.24\,\ep^4
+ 20.75\,\ep^5
\ . 
\end{array}
\eeq
Note that $\eta^{(5)}$ has decreased by about 30 \% in
comparison with the (incorrect) results of Ref.~\cite{ep5}, $\nu ^{(5)}$
has increased by about 10 \%, and $\omega^{(5)}$  increased by about
0.6 \% in comparison with \cite{ep5}.

It is known that the series of the
$\ep$-expansion 
are asymptotic series and special resummation
techniques
\cite{asymp,jkbr} should be applied to obtain reliable estimates
of the critical exponents. Although the size
of the $\ep^5$ terms in the physical dimension (i.e., at   
$\ep=1$) is very large, their contribution to the exponents 
in the {\em resummed\/} series is very small.
This is why even large relative
changes of the
$\ep^5$ coefficients 
turn out not to
change
the critical exponents~\cite{jkl} within the accuracy
of previous determinations~\cite{ep5,gz}.


\begin{thebibliography}{999999}
\bibitem{jkl}
This was checked in detail 
by W. Janke and H. Kleinert using the resummation programs of Ref. \cite{jkl}.
Since the deviations from the previous numbers were 
found to be very small the results 
were not published.
\end{thebibliography}
\end{document}